\documentclass[%
 reprint,
 amsmath,amssymb,
 aps]{revtex4-1}

\usepackage{graphicx}
\usepackage{dcolumn}
\usepackage{bm}
\usepackage{hyperref}
\usepackage{braket}
\usepackage{color}
\usepackage{colortbl}

\begin{document}

\preprint{APS/123-QED}

\title{Quasinormal Modes and Hawking-Unruh effect in Quantum Hall Systems: Lessons from Black Hole Phenomena}

\author{Suraj S Hegde}
\author{Varsha Subramanyan} 
\author{Barry Bradlyn}
\author{Smitha Vishveshwara}
\affiliation{Department of Physics and Institute for Condensed Matter Theory, University of Illinois at Urbana-Champaign, Urbana, IL, 61801-3080, USA
}%



\begin{abstract}
 In this work, we propose the quantum Hall system as a platform for exploring black hole phenomena. By exhibiting deep rooted commonalities between lowest Landau level and spacetime symmetries, we show that features of both quantum Hall and gravitational systems can be elegantly captured by a simple quantum mechanical model, the inverted harmonic oscillator. Through this correspondence, we argue that radiation phenomena in gravitational situations, such as presented by W.~G.~Unruh and S.~Hawking, bears a parallel with saddle-potential scattering of quantum Hall quasiparticles. We also find that scattering by the quantum Hall saddle potential can mimic the signature quasinormal modes in black holes, such as theoretically demonstrated through Gaussian scattering off a Schwarzschild black hole by C.~V.~Vishveshwara. We propose a realistic quantum Hall point contact setup for probing these temporally decaying modes in quasiparticle tunneling, offering a new mesoscopic parallel for black hole ringdown.

\end{abstract}

\maketitle



General relativity and quantum mechanics are the two cornerstones of modern physics, the former describing physics at the astronomical scales and the latter primarily at micro- and mesoscopic scales. 
Outside of quantum gravity, these fields of studies are largely viewed as disjoint, each with a host of exotic physical phenomena. 
Recently, numerous works have drawn analogies between these disparate scales. 
Quantum systems in fact offer a multitude of ways to probe relativistic phenomena, from mimicking curved spacetimes \cite{Barcelo} to investigating dynamic geometric backgrounds \cite{Hartnoll, Haldane, Liu, Lapa, Gromov}.
In this Letter, 
we show that signature features of black holes, and scattering in quantum Hall systems can be remarkably unified by a mapping to single particle physics in the presence of an inverted harmonic oscillator (IHO) potential. The IHO model exhibits potential scattering and temporally decaying modes, features that have made the model invaluable in a broad variety of contexts since the birth of quantum mechanics \cite{kemble1937,Landau,Barton86,Yuce}. From its infancy, phenomena such as particle decay \cite{Gamow} and  metastability \cite{Boyanovsky} have been analyzed using the IHO. In developments across the decades, the IHO has played key roles in the context of many disparate fields of physics and mathematics\cite{Berry, Sierra, Bhaduri, Miller, Cuchietti, Baskoutas,Gentilini,Friess, Maldacena05,
Cremonini, Schoutens,Guth}. Thus, through its powerful simplicity, the IHO serves as an archetype for phenomena in numerous realms, much like the more familiar simple harmonic oscillator. Here, the IHO plays a crucial role in demonstrating how the confluence of black hole and quantum Hall physics can transfer insights between the astronomical and the mesoscopic realms.

 In this Letter, we demonstrate that the physics of quantum Hall systems in strained or  point contact geometries makes manifest two deep mathematical structures underlying  black hole phenomena. Specifically, the quantum Hall situation describes lowest Landau level (LLL) physics in the presence of a saddle potential. (Quasi)-particles moving in this potential are squeezed as a function of time; this operation of LLL squeezing is formally equivalent to time-evolution near the event horizon outside a black hole\cite{Sewell82}. Hence, we show that the transmission probability near the point contact region manifest in conductance measurements naturally follows the thermal form associated with Hawking radiation. The second mathematical structure captured by such point contact physics is the scattering off the effective potential that lies outside a black hole event horizon. Quasi-normal modes, a hallmark temporally decaying feature of this scattering, are the fingerprint signatures of black holes in gravitational wave detection and have been hinted at by the Laser Interferometer Gravitational Wave Observatory (LIGO) discovery. Inspired by this black hole phenomenon, here we highlight the presence of analogous temporally decaying modes in quantum Hall tunneling and propose a hitherto unexplored time-resolved measurement in quintessential mesoscopic setting of quantum point contacts. 
 

In what follows, we demonstrate how the semi-classical black hole scattering problem can be reduced to one-dimensional scattering off an IHO potential. We then show that the non-commutative structure of the lowest-Landau level gives rise to the same scattering description in the presence of a saddle potential. Distinct from the black hole scattering context, we show that Hamiltonian describing the IHO is isomorphic to the Rindler Hamiltonian associated with time-evolution near the event horizon. This symmetry-based link puts the previously observed connection between quantum Hall transmission probabilities and black hole thermality\cite{Stone12} on a stronger footing than a cursory analogy. Finally, we  perform an in-depth analysis of quasinormal mode decay as an unexplored probe of quantum Hall dynamics.  


{{\bf \it Black Hole Scattering:} }
A black hole is formed by the gravitational collapse of a massive object. It is characterized by an event horizon which acts as a one-way membrane in spacetime: matter can enter the horizon, but cannot leave. Several classical gravity and holographic phenomena , such as black hole collisions and quasinormal modes, are described through wave scattering in the black hole spacetime \cite{Chandrasekhar,Konoplaya11}. Here we focus on such scattering for a  massless scalar field $\Phi$ to provide a simple, illustrative derivation of IHO dynamics. The equation of motion for $\Phi$ in the background metric $g_{\mu \nu}$ is given by \cite{Chandrasekhar,MTW}
 \begin{equation}
 \frac{1}{\sqrt{-g}}\frac{\partial}{\partial x^{\mu}} \bigg( g^{\mu \nu} \sqrt{-g} \frac{\partial}{\partial x^{\nu}} \Phi \bigg) =0.
 \label{Laplacian}
 \end{equation}
 
\begin{figure}[t]
\includegraphics[width=0.4\textwidth]{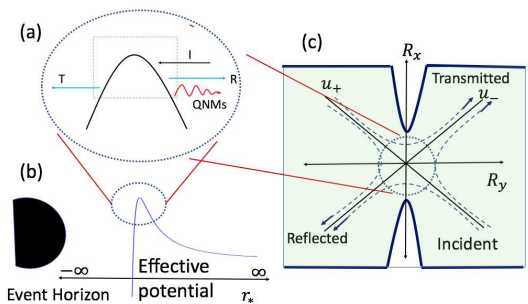}
\caption{ a) Schematic of the inverted harmonic oscillator potential and its scattering features of i) regular transmission (T) and reflection (R) of an incident state (I) and ii) temporally decaying quasinormal modes (QNMs), as applicable to two physically distinct phenomena (b and c):  b) Dynamics of a field in a black hole spacetime results in i) Hawking-Unruh radiation and ii) signature QNMs associated with black hole ringdown. c) Scattering in the quantum Hall lowest Landau level via a saddle potential leads to i) quasiparticle tunneling and, ii) as we predict, QNMs which could be observed for dilute enough quasiparticle sources.}\label{Scatter}
\end{figure} 
 
For the Schwarzschild metric associated with an uncharged, non-rotating black hole whose event horizon is at $r=2GM/c^2$, we can decompose the field into its radial and angular parts. For each angular momentum component $\ell$, the scalar field experiences an effective potential with its maximum outside the event horizon. The equation for the radial component $\Psi(r)$ reduces to the well-known form\cite{ReggeWheeler,Andersson,Chandrasekhar,Vishveshwara,Schutz85}

\begin{equation}
\frac{\partial^2 \Psi }{\partial r_*^2} + V(\ell, r_*) \Psi =0.
\label{WaveScatter}
\end{equation} 

Here $r_*$ is the ``tortoise co-ordinate''\cite{MTW} in which the event horizon is at negative infinity and $V(\ell, r_*)$ is the effective potential. Higher spin fields, such as for gravitational waves, can be treated analogously. Crucially, the potential has a maximum  $V_0=V(\ell, r_0)$ at a point $r_*=r_0$, and can be approximated by an inverted harmonic potential nearby. Thus, in the semiclassical approximation,  Eq.~(\ref{WaveScatter}) takes the Weber form
\cite{Schutz85,Ford59} :
\begin{equation}
-\frac{d^2 \Psi}{d\tilde{r}_*^2}- \frac{1}{4}\tilde{r}_*^2 \Psi  =(\nu +\frac{1}{2})\Psi. \label{eq:weber}
\end{equation}
Here,  $\tilde{r}*= (2V_0 '')^{1/4}(r_*-r_0)$ measures the dimensionless distance from the maximum. We identify $\nu+1/2 = -V_0 /(2V_0 '')^{1/2}$ so as to cast Eq.~(\ref{eq:weber}) in the standard form. This derivation highlights the relevance of the IHO in black hole scattering~\cite{Schutz85,Konoplaya11, Mashhoon, Andersson}. In fact, the IHO captures essential dynamics even in the presence of gravitational backreaction~\cite{Betzios16, Dray85} and the semiclassical limit of scattering off any smooth barrier \cite{Ford59}.

{{\bf \it Quantum Hall Scattering:}}  We now show how the same scattering problem in Eq.~(\ref{eq:weber}) arises in a quantum Hall system in the presence of a saddle potential. We consider a two-dimensional system of electrons confined to the $x-y$ plane in the presence of a magnetic field $B\hat{\mathbf{z}}=\nabla\times\vec{A}$. Neglecting interactions, as is appropriate for an integer quantum Hall state, the Hamiltonian takes the form
\begin{align}
H_0=\frac{1}{2m}|\vec{\pi}|^2+ \lambda(x^2-y^2),
\label{QHam}
\end{align}
 where $\hat{\pi_i}=\hat{p_i}-e\hat{A}_i/c$ is the kinetic momentum, and the index $i=x,y$ ranges over the spatial directions. The parameter $\lambda$ captures the strength of the saddle potential. This potential models bulk quasiparticle tunneling through a point contact\cite{Fertig87,Stone12, Buttiker90,Matthews} and can serve as an anyon beam-splitter \cite{Smitha10}. 
 
 We introduce the guiding center operators $\hat{R_i}=\hat{x_i}+\frac{\ell_B^2}{\hbar}\epsilon_{ij}\hat{\pi_j},$ where $\ell_B=\sqrt{\hbar c/(eB)}$ is the magnetic length. 
  Upon projection to the lowest Landau level, the dynamics is determined entirely by the projected potential, which is a function of the $R_i$ only\cite{girvinjach}. The guiding center coordinates form a non-commutative plane, obeying $[\hat{R_i},\hat{R_j}]=-il_B^2\epsilon_{ij}$. This commutation rule yields an effective ``Planck's constant'' $\hbar_{eff}=l_B^2$ in this phase space that can be tuned through the applied magnetic field to access the semi-classical limit. Identifying $\hat{R_x}/(\sqrt{2}\ell_B)$ as a ``momentum" $\tilde{P}$, and $\sqrt{2}\hat{R_y}/\ell_B$ as a ``position" $\tilde{X}$, the projected Hamiltonian is that of the IHO:
   \begin{equation}
 H_{IHO}= 
 (2\hbar_{eff}\lambda) (\tilde{P}^2-\frac{1}{4}\tilde{X}^2).\label{eq:projham}
 \end{equation}
Identifying $\tilde{X}$ with $\tilde{r}_*$, and the eigenenergy $E_\nu=2\hbar_{eff}\lambda(\nu+\frac{1}{2})$ for any real $\nu$, the Schr\"{o}dinger equation for this Hamiltonian takes the form of Eq.~(\ref{eq:weber}).
Note that even without LLL projection, dynamics in the presence of a saddle potential decouples into intra- and inter-Landau level components; the intra-Landau level dynamics is also governed by the applied potential\cite{Fertig87}.

 The saddle potential can also be generated from area preserving shear deformations of two-dimensional electrons in a magnetic field. The generators $\hat{J}_{ij}$ of these transformations are\cite{ReadVisc,Bradlyn2012}
\begin{equation}
\hat{J}_{ij}=-\frac{1}{2}\{\hat{x}_i,\hat{\pi}_j\}+\frac{1}{4} \{\hat{x}_k,\hat{\pi}_k\} \delta_{ij}+\frac{\hbar}{2\ell_B^2}\epsilon_{ij}\hat{x}_j \hat{x}_k. \label{gen}
\end{equation}
 The generators satisfy the $\mathfrak{sl}(2,\mathbb{R})$ commutation relations $i[J_{ij},J_{kl}]=\delta_{il}J_{kj}-\delta_{jk}J_{il}$. These deformations are important in the study of Hall viscosity\cite{ReadVisc}. Of the three independent generators of the Lie algebra, the shear generator 
 
 \begin{equation} 
 V= \frac{2\lambda\ell_B^2}{\hbar}(J_{xy}+J_{yx})=\lambda(R_x^2 -R_y^2)/4 + \frac{\lambda\ell_B^4}{\hbar^2} (\pi_y^2 - \pi_x^2)
 \end{equation}
 is of interest. Upon projecting it to the lowest Landau level, the kinetic momenta can be neglected due to vanishing matrix elements between any states in the same Landau level, leaving $H_{IHO}$ of Eq.~(\ref{eq:projham}). The Hamiltonian $H_{IHO}$ is in fact a squeezing operator\cite{yuensqueezing}, which generates Bogoliubov transformations of the LLL eigenstates\cite{Smitha10,Stone12}. 

{\bf {\it The Hawking-Unruh effect-}}
Given the common IHO starting point of Eq.~(\ref{eq:projham}), we now show that scattering off such a barrier captures both quantum Hall tunneling and radiation in black holes. 
To evaluate the scattering properties, we canonically transform to the basis $\hat{u}^{\pm} =\frac{\hat{P} \pm \hat{X}}{\sqrt{2}}$ \cite{Maldacena05}, where $\hat{u}^{\pm}$ are the incoming and outgoing states, as shown in Fig.~(\ref{Scatter}). In this basis, $H_{IHO}$ takes the form
\begin{equation}
H= \mp i2\hbar_{eff}\lambda \bigg( u^{\pm} \partial_{u^{\pm}} + \frac{1}{2}\bigg)\label{dilop},
\end{equation}
which is the generator of rescalings, equivalently dilatations, of $u^\pm$\cite{FrancescoCFT}. The eigenstates are of the form $\frac{1}{\sqrt{2 \pi}}(\pm u^\pm)^{iE-1/2} \Theta(\pm u^\pm)\textnormal{ ,where }\Theta(u^{\pm})$ is the Heaviside function and $E=\epsilon/2\hbar_{eff}\lambda$. The expansion in the eigenbasis of the dilatation operator is a Mellin transform \cite{Moses72, Cycon87}, just as the Fourier transform is associated with the translation generator. 
The Mellin transform connects the incoming and outgoing state, thus yielding the scattering matrix (S-matrix). The S-matrix has the form\cite{kemble1937,Maldacena05,Betzios16}
\begin{equation}
  \hat{\mathcal{S}}= \frac{1}{\sqrt{2 \pi}} \Gamma\bigg( \frac{1}{2}-iE  \bigg)  \left(\begin{array}{cc}
 e^{i\pi/4}e^{\pi E/2}&
  e^{-i\pi/4}e^{-\pi E/2}\\
   e^{-i\pi/4}e^{-\pi E/2}& e^{i\pi/4}e^{\pi E/2}\end{array}\right), \label{smat}
  \end{equation} 
  where $\Gamma$ is the Gamma function. 
  
  The transmission probability obtained from  this $S-$matrix yields the suggestive thermal form
  \begin{equation}
  |t|^2 = \frac{1}{1+e^{\beta \epsilon}},\;\;\textnormal{where}\;\; \beta=\frac{\pi}{\hbar_{eff}\lambda}. \label{dist}
  \end{equation}
In the quantum Hall case, the transmission probability describes tunneling of bulk quasiparticles between equipotential contours of the saddle potential. Such tunneling is ubiquitous in the quantum Hall system, for instance in scattering between peaks or valleys in a disordered bulk potential landscape \cite{Chalker,Huckestein}. The tunneling form is also applicable for tunneling between edge channels across a pinched Hall bar, providing a derivation for the transmission probability via the bulk \cite{Fertig87, Sen}.

In the case of black holes, Eq.~(\ref{dist}) can be interpreted as the thermal distribution associated with the Hawking-Unruh effect. While there are many ways to understand the effect \cite{Parikh00, Vanzo11, Sewell82,Bisognano,Robinson05}, we closely follow the treatment of Refs.~\cite{Hawking,Hawking75,Unruh76,Stone12,volovik1999vierbein}. To recapitulate, the thermal distribution stems from comparing vacua associated with inertial and accelerating observers. Specifically, we consider the vacuum state of a fermionic field (satisfying Eq.~(\ref{Laplacian})) in a Minkowski (inertial) spacetime $(t,x)$. A uniformly accelerating observer characterized by ``Rindler co-ordinates'' $(\tau,\xi)$, where $t= e^{\xi} \sinh(\tau), x=e^{\xi} \cosh(\tau)$, observes a different vacuum. Note that translations of $\tau$, generated by the ``Rindler Hamiltonian'' correspond to Lorentz boosts in $(t,x)$. In Minkowski spacetime, we can mode-expand the field in the basis of plane waves $e^{i\omega v}\textnormal{ (}v=x-t$ is the ``light-cone'' coordinate).The expansion of the fields in terms of the Rindler space is then in the  eigenbasis $e^{i\Omega (\tau -\xi)}=v^{i\Omega}$. The Lorentz boost generator relates the modes in the two frames; a calculation shows that the fields are related by a Bogoliubov transformation\cite{Mukhanov,Crispino08}. Due to the Bogoliubov transformation, the Minkowski vacuum is perceived as a thermal distribution, such as depicted in Eq.~(\ref{dist}), for the Rindler observer. In the presence of a black hole, such an analysis results in the Hawking radiation seen by an asymptotic observer\cite{Hawking, Hawking75}.

 It might be surprising that the two disparate platforms--one a relativistic spacetime and the other a quantum Hall system,--exhibit such similarities. The connections are far from cursory and emerge from the underlying symmetry transformations: first consider the $2+1$-dimensional Minkowski spacetime $(t,\vec{x})$ with the metric $ds^2=dt^2-d \vec{x}^2$. The Lorentz boost is one of the generators of the full spacetime symmetry algebra $\mathfrak{so}(2,1)$. In  the quantum Hall system, the projected saddle potential (or the IHO/ dilatation operator) is one of the generators of transformations that preserve the commutator $[R_x,R_y]=-il_B^2$. The generators respect a $\mathfrak{sp}(2,\mathbb{R})$ algebra. These two Lie algebras are isomorphic, and coincide with that of area/flux preserving deformations given below Eq.~(\ref{gen}). Similarly, we can relate quantum states in the two situations as well. Note that the Rindler Hamiltonian and the IHO/dilatation operator in the LLL  share similar eigenfunctions, upon identifying $\Omega=E-i/2$, where the extra $1/2$ is a result of the noncommutativity of position and momentum in the quantum Hall case. In the relativistic derivation, this extra $1/2$ arises from the boost acting on the spin component of the fermionic fields\cite{Alsing}. As recently detailed\cite{Arzano}, these modes determine the Bogoliubov coefficients through the Mellin transform and lead to Eq.~(\ref{dist}) in both cases. This parallel between the quantum Hall transmission probability and the Hawking-Unruh effect was first explored for edge states in Ref.~\cite{Stone12}; here, we have generalized the treatment to \emph{bulk} quasiparticle scattering. 

 We emphasize that there is a  distinction in the way the IHO relates to the two black hole phenomena considered in this work. The Hawking-Unruh effect explicitly involves second quantization and quantum effects across the event horizon, and the IHO appears here as the Rindler Hamiltonian corresponding to boosts. In contrast, black hole scattering concerns IHO potentials outside the event horizon. Moreover, the associated QNMs that we now turn to, are present in both quantum and classical situations. 
 


{\bf {\it Quasinormal modes (QNMs): }} QNM decay was originally predicted in the context of the stability of Schwarzschild black holes by C. V. Vishveshwara in Ref.~\cite{Vishveshwara},  as damped outgoing oscillations in response to incoming Gaussian wave packets. This phenomenon has come to be a central point of study in black hole physics \cite{Press,Kokkotas,Konoplaya11,Berti, Hubeny, Panotopoulos,Teukolsky,ChandraDetweiler}. Here, our map of black hole scattering to the quantum IHO problem \cite{Mashhoon, Schutz85} in Eq.~(\ref{eq:weber}) and the resultant S-matrix Eq.~(\ref{smat}) enable us to explicitly characterize these damped oscillations, even if to the simplest approximation. The map also connects the black hole QNMs to previously unexplored dynamics of quantum Hall systems in a saddle potential.

\begin{figure}[t]
\includegraphics[width=0.4\textwidth]{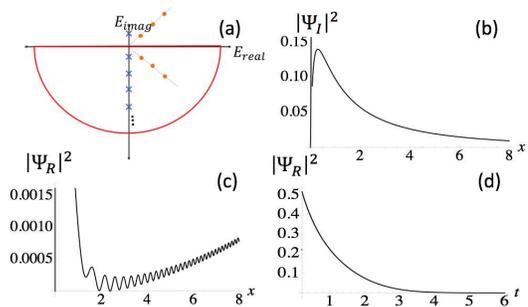}
\caption{ Plots of wave-packet scattering  off the IHO showing quasinormal mode behavior, as obtained form analytical calculations.(a) Pole structure of the S-matrix of the IHO in the complex energy plane. The blue crosses indicate the resonant poles of the IHO, and the purple boxes indicate generic resonance poles for some arbitrary potential barrier. Closing the contour in the lower half plane is determined by the fact that we have picked outgoing boundary conditions (b) A wave-packet composed of scattering energy  eigenstates, impinging on the barrier from the right. (c) The scattered wavepacket shows the ``quasinormal ringdown"; the form takes into account only a single pole for illustrative purposes. The scattered state escapes to infinity as seen in its finite amplitude at large x, but as shown in
(d),  it exhibits an exponential time decay for a given point x after $t>\tau \log |x/\ell|$ }\label{qnm1}
\end{figure}

The S-matrix of Eq.~(\ref{smat}) directly accesses the QNMs in its pole structure when analytically extended to the complex energy plane\cite{Perelomov}. Specifically, the residues of the poles are resonant (quasi-stationary, or Gamow) states\cite{Parravacini,Bohm}. These states explicitly decay in time and thus lie outside the standard Hilbert space \cite{Chruscinski03,Chruscinski04}. Note that this decay does not require any non-unitary time evolution, rather it is a consequence of the (purely outgoing) boundary conditions on the states. This can be compared with dissipation in the Landauer-Buttiker formalism for conductivity, where decay comes as a consequence of the boundary condition on states in the reservoir\cite{Datta}. The S-matrix for the IHO in Eq.~(\ref{smat}) has poles at $E_n=-i(n+\frac{1}{2}); n=0,1,2..$,  coming from the Gamma function (see  Fig.~(\ref{qnm1})). In order to probe these poles, as in both quantum scattering\cite{Perelomov} and black hole perturbations\cite{Konoplaya11, Andersson},  we employ  a \emph{dynamic} scattering process of impinging wavepackets onto the potential, rather than energy eigenfunctions.

A typical incident wavepacket, as shown in Fig.~(\ref{qnm1}), is composed of a superposition of energy states: $|\psi_{in}\rangle = \int d\epsilon\;f(\epsilon)|\epsilon\rangle$. For simplicity, we take $f(\epsilon)$ to be a Gaussian centered at an energy $\epsilon_*$ with width $\delta_\epsilon$. Importantly, the effect of the resonant QNM poles becomes manifest in the reflected wave at position $x$ only for time $t>\tau \log |x/\ell|$, where $\tau=(2\hbar_{eff}\lambda)^{-1}$ and $\ell$ is the oscillator length. This can be understood from the classical trajectory of the model, $x \sim e^{t/\tau}$. For widths $\delta_\epsilon\ll \hbar_{eff}\lambda$, the decay is governed by the residue of a single pole. For example, the reflected QNM decay due to the $n=0$ pole has the spatio-temporal form 
 \begin{equation}
\Psi_{r0} \sim e^{-t/\tau} e^{\log|\sqrt{2}x/l|}.
\label{eq:QNMDecay}
\end{equation}
This shows that QNM decay gives the asymptotic description of the change in probability as a scattered wavepacket passes a (fixed) distant point.

 The scattered packet also undergoes oscillations, as depicted in Fig.~(\ref{qnm1}(c)). For any fixed position, Fig.~(\ref{qnm1}(d)) depicts the characteristic temporal decay. These plots only take into account the pole at $n=0$. 
Although we study resonances here as arising from the poles of the S-matrix a wave-packet scattering, they can be equivalently cast as states in a Rigged Hilbert space having complex eigen-energies \cite{Chruscinski03,Chruscinski04, Parravacini,Bohm,Shimbori00} arising in an open system with purely outgoing boundary conditions. It is these boundary conditions that allow for seemingly nonunitary decay. Such QNM decay arises in any system having a potential landscape characterized by a local maxima, such as in Gamow's theory of radioactivity\cite{Gamow}. Thus, this analysis predicts the existence of black hole-type QNMs in our quantum Hall setting.
 
{\it QNMs in physical observables:} A direct measure of QNM decay would require a time-resolved non-equilibrium setting. In the black hole situation, the LIGO breakthrough recorded ringdown signals in cataclysmic black hole mergers, ushering in the era of multi-messenger astronomy \cite{Isi19}. For a Schwarzschild black hole of one solar mass, the calculated decay time is 0.35ms \cite{Chirenti}; the decay time in the remarkable first measurement by LIGO  from binary black hole merger was 4 milliseconds \cite{LIGO1}. 

We now propose a setup in the quantum Hall situation for observing QNMs and  derive analogous estimates. As shown in Fig.~(\ref{Scatter}), a pinched point contact geometry creates a saddle potential in the bulk. Sources of bulk or edge state quasiholes undergo saddle potential scattering and tunneling as described above. An indirect measure of QNM poles in the scattering matrix would be a Lorentzian form for the associated tunneling conductance, known as the Breit-Wigner distribution\cite{McEuen}. A direct measure would require a dilute beam of incoming quasiholes \cite{Heiblum} that could enable time resolving QNMs in the out-going beam. Considering a point contact of width $d$, for applied voltage $V$ and a background magnetic field $B$, we have that $\tau=d^2B/2V$, and $\lambda=eV/d^2$. Putting in typical numbers for a split-gate graphene junction \cite{Floser}, we find that $\tau\sim 400ps$, while $\hbar_{eff}\lambda\sim 125mK \ll \hbar\omega_c$, and $\ell_B\sim 5nm$. We anticipate then that the ringdown associated with quasihole wavepackets should be observable at a distance of order $100nm$ from the point contact at times on the order of nanoseconds. 
Finally, we note that realistic scattering potentials would have finite range, unlike the unbounded IHO; the P\"{o}schl-Teller potential yields a tractable candidate for such analyses \cite{Aguirregabiria, Buttiker90}. We suggest that the actual recorded temporal decay in a given measurement would serve to recreate the underlying potential.

In conclusion, we have demonstrated that black hole and quantum Hall scattering dynamics naturally fall under a common umbrella dictated by properties of the inverted harmonic oscillator. These parallels not only draw connections between well known effects, they offer ground for unearthing new phenomena in both realms, such as QNMs in quantum Hall systems. We expect non-Poisson corrections to the current correlation function in quantum Hall noise experiments to provide an additional window into these temporally decaying modes. We hope that these parallels allow for future work to employ the quantum Hall system for even studying black holes and IHO physics in the context of recent connections to quantum chaos\cite{Maldacena16,Shenker14,Bhaduri, Hashimoto, Morita18,Berry,Sierra}. Our symmetry-based arguments also suggest several other experimentally accessible realizations of relativistic kinematics in quantum Hall system, such as Thomas precession. Thus, in drawing specific parallels between gravitational and quantum Hall physics, and transferring lessons from black hole scattering to new predictions in point contact geometries, we hope to have planted seeds of inspiration for fertile exchange of ideas between the two realms.  


\begin{acknowledgments}
We are grateful to N. Andersson and M. Stone for their caring and insightful comments. We dedicate this work to black hole physicist, mentor, and father, C. V. Vishveshwara (1938-2017), with whom we shared the marvels of quantum Hall physics and who witnessed LIGO's ringdown observations in his lifetime decades after his prediction of black hole quasinormal modes.
\end{acknowledgments}

\bibliography{BH-QHnotes}

\begin{thebibliography}{88}%
\makeatletter
\providecommand \@ifxundefined [1]{%
 \@ifx{#1\undefined}
}%
\providecommand \@ifnum [1]{%
 \ifnum #1\expandafter \@firstoftwo
 \else \expandafter \@secondoftwo
 \fi
}%
\providecommand \@ifx [1]{%
 \ifx #1\expandafter \@firstoftwo
 \else \expandafter \@secondoftwo
 \fi
}%
\providecommand \natexlab [1]{#1}%
\providecommand \enquote  [1]{``#1''}%
\providecommand \bibnamefont  [1]{#1}%
\providecommand \bibfnamefont [1]{#1}%
\providecommand \citenamefont [1]{#1}%
\providecommand \href@noop [0]{\@secondoftwo}%
\providecommand \href [0]{\begingroup \@sanitize@url \@href}%
\providecommand \@href[1]{\@@startlink{#1}\@@href}%
\providecommand \@@href[1]{\endgroup#1\@@endlink}%
\providecommand \@sanitize@url [0]{\catcode `\\12\catcode `\$12\catcode
  `\&12\catcode `\#12\catcode `\^12\catcode `\_12\catcode `\%12\relax}%
\providecommand \@@startlink[1]{}%
\providecommand \@@endlink[0]{}%
\providecommand \url  [0]{\begingroup\@sanitize@url \@url }%
\providecommand \@url [1]{\endgroup\@href {#1}{\urlprefix }}%
\providecommand \urlprefix  [0]{URL }%
\providecommand \Eprint [0]{\href }%
\providecommand \doibase [0]{http://dx.doi.org/}%
\providecommand \selectlanguage [0]{\@gobble}%
\providecommand \bibinfo  [0]{\@secondoftwo}%
\providecommand \bibfield  [0]{\@secondoftwo}%
\providecommand \translation [1]{[#1]}%
\providecommand \BibitemOpen [0]{}%
\providecommand \bibitemStop [0]{}%
\providecommand \bibitemNoStop [0]{.\EOS\space}%
\providecommand \EOS [0]{\spacefactor3000\relax}%
\providecommand \BibitemShut  [1]{\csname bibitem#1\endcsname}%
\let\auto@bib@innerbib\@empty
\bibitem [{\citenamefont {Barcel\'o}\ \emph {et~al.}(2005)\citenamefont
  {Barcel\'o}, \citenamefont {Liberati},\ and\ \citenamefont
  {Visser}}]{Barcelo}%
  \BibitemOpen
  \bibfield  {author} {\bibinfo {author} {\bibfnamefont {C.}~\bibnamefont
  {Barcel\'o}}, \bibinfo {author} {\bibfnamefont {S.}~\bibnamefont {Liberati}},
  \ and\ \bibinfo {author} {\bibfnamefont {M.}~\bibnamefont {Visser}},\ }\href
  {\doibase 10.12942/lrr-2005-12} {\bibfield  {journal} {\bibinfo  {journal}
  {Living Reviews in Relativity}\ }\textbf {\bibinfo {volume} {8}},\ \bibinfo
  {eid} {12} (\bibinfo {year} {2005})},\ \Eprint
  {http://arxiv.org/abs/gr-qc/0505065} {gr-qc/0505065} \BibitemShut {NoStop}%
\bibitem [{\citenamefont {Hartnoll}\ \emph {et~al.}(2018)\citenamefont
  {Hartnoll}, \citenamefont {Lucas},\ and\ \citenamefont {Sachdev}}]{Hartnoll}%
  \BibitemOpen
  \bibfield  {author} {\bibinfo {author} {\bibfnamefont {S.~A.}\ \bibnamefont
  {Hartnoll}}, \bibinfo {author} {\bibfnamefont {A.}~\bibnamefont {Lucas}}, \
  and\ \bibinfo {author} {\bibfnamefont {S.}~\bibnamefont {Sachdev}},\
  }\href@noop {} {\emph {\bibinfo {title} {Holographic quantum matter}}}\
  (\bibinfo  {publisher} {MIT press},\ \bibinfo {year} {2018})\BibitemShut
  {NoStop}%
\bibitem [{\citenamefont {Haldane}(2011)}]{Haldane}%
  \BibitemOpen
  \bibfield  {author} {\bibinfo {author} {\bibfnamefont {F.~D.~M.}\
  \bibnamefont {Haldane}},\ }\href {\doibase 10.1103/PhysRevLett.107.116801}
  {\bibfield  {journal} {\bibinfo  {journal} {Phys. Rev. Lett.}\ }\textbf
  {\bibinfo {volume} {107}},\ \bibinfo {pages} {116801} (\bibinfo {year}
  {2011})}\BibitemShut {NoStop}%
\bibitem [{\citenamefont {Liu}\ \emph {et~al.}(2018)\citenamefont {Liu},
  \citenamefont {Gromov},\ and\ \citenamefont {Papi\ifmmode~\acute{c}\else
  \'{c}\fi{}}}]{Liu}%
  \BibitemOpen
  \bibfield  {author} {\bibinfo {author} {\bibfnamefont {Z.}~\bibnamefont
  {Liu}}, \bibinfo {author} {\bibfnamefont {A.}~\bibnamefont {Gromov}}, \ and\
  \bibinfo {author} {\bibfnamefont {Z.}~\bibnamefont
  {Papi\ifmmode~\acute{c}\else \'{c}\fi{}}},\ }\href {\doibase
  10.1103/PhysRevB.98.155140} {\bibfield  {journal} {\bibinfo  {journal} {Phys.
  Rev. B}\ }\textbf {\bibinfo {volume} {98}},\ \bibinfo {pages} {155140}
  (\bibinfo {year} {2018})}\BibitemShut {NoStop}%
\bibitem [{\citenamefont {Lapa}\ \emph {et~al.}(2018)\citenamefont {Lapa},
  \citenamefont {Gromov},\ and\ \citenamefont {Hughes}}]{Lapa}%
  \BibitemOpen
  \bibfield  {author} {\bibinfo {author} {\bibfnamefont {M.~F.}\ \bibnamefont
  {Lapa}}, \bibinfo {author} {\bibfnamefont {A.}~\bibnamefont {Gromov}}, \ and\
  \bibinfo {author} {\bibfnamefont {T.~L.}\ \bibnamefont {Hughes}},\
  }\href@noop {} {\  (\bibinfo {year} {2018})},\ \Eprint
  {http://arxiv.org/abs/1809.06386 [cond-mat.str-el]} {arXiv:1809.06386
  [cond-mat.str-el]} \BibitemShut {NoStop}%
\bibitem [{\citenamefont {Gromov}\ and\ \citenamefont {Son}(2017)}]{Gromov}%
  \BibitemOpen
  \bibfield  {author} {\bibinfo {author} {\bibfnamefont {A.}~\bibnamefont
  {Gromov}}\ and\ \bibinfo {author} {\bibfnamefont {D.~T.}\ \bibnamefont
  {Son}},\ }\href {\doibase 10.1103/PhysRevX.7.041032} {\bibfield  {journal}
  {\bibinfo  {journal} {Phys. Rev. X}\ }\textbf {\bibinfo {volume} {7}},\
  \bibinfo {pages} {041032} (\bibinfo {year} {2017})}\BibitemShut {NoStop}%
\bibitem [{\citenamefont {Kemble}(1937)}]{kemble1937}%
  \BibitemOpen
  \bibfield  {author} {\bibinfo {author} {\bibfnamefont {E.~C.}\ \bibnamefont
  {Kemble}},\ }\href@noop {} {\emph {\bibinfo {title} {The Fundamental
  Principles of Quantum Mechanics}}}\ (\bibinfo  {publisher} {McGraw-Hill},\
  \bibinfo {year} {1937})\BibitemShut {NoStop}%
\bibitem [{\citenamefont {Landau}\ and\ \citenamefont {Lifshitz}()}]{Landau}%
  \BibitemOpen
  \bibfield  {author} {\bibinfo {author} {\bibfnamefont {L.~D.}\ \bibnamefont
  {Landau}}\ and\ \bibinfo {author} {\bibfnamefont {E.~M.}\ \bibnamefont
  {Lifshitz}},\ }\href@noop {} {\emph {\bibinfo {title} {Quantum
  Mechanics(Non-relativistic theory): Course on theoretical physics vol.3}}}\
  (\bibinfo  {publisher} {Elsevier Butterworth-Heinemann})\BibitemShut
  {NoStop}%
\bibitem [{\citenamefont {Barton}(1986)}]{Barton86}%
  \BibitemOpen
  \bibfield  {author} {\bibinfo {author} {\bibfnamefont {G.}~\bibnamefont
  {Barton}},\ }\href@noop {} {\bibfield  {journal} {\bibinfo  {journal} {Annals
  of Physics}\ }\textbf {\bibinfo {volume} {166}},\ \bibinfo {pages} {322}
  (\bibinfo {year} {1986})}\BibitemShut {NoStop}%
\bibitem [{\citenamefont {Yuce}\ \emph {et~al.}(2006)\citenamefont {Yuce},
  \citenamefont {Kilic},\ and\ \citenamefont {Coruh}}]{Yuce}%
  \BibitemOpen
  \bibfield  {author} {\bibinfo {author} {\bibfnamefont {C.}~\bibnamefont
  {Yuce}}, \bibinfo {author} {\bibfnamefont {A.}~\bibnamefont {Kilic}}, \ and\
  \bibinfo {author} {\bibfnamefont {A.}~\bibnamefont {Coruh}},\ }\href
  {http://stacks.iop.org/1402-4896/74/i=1/a=014} {\bibfield  {journal}
  {\bibinfo  {journal} {Physica Scripta}\ }\textbf {\bibinfo {volume} {74}},\
  \bibinfo {pages} {114} (\bibinfo {year} {2006})}\BibitemShut {NoStop}%
\bibitem [{\citenamefont {Gamow}(1928)}]{Gamow}%
  \BibitemOpen
  \bibfield  {author} {\bibinfo {author} {\bibfnamefont {G.}~\bibnamefont
  {Gamow}},\ }\href@noop {} {\bibfield  {journal} {\bibinfo  {journal}
  {Zeitshrift f¨ur Physik}\ }\textbf {\bibinfo {volume} {51}},\ \bibinfo
  {pages} {204} (\bibinfo {year} {1928})}\BibitemShut {NoStop}%
\bibitem [{\citenamefont {Boyanovsky}\ \emph {et~al.}(1995)\citenamefont
  {Boyanovsky}, \citenamefont {Holman}, \citenamefont {Lee}, \citenamefont
  {Silva},\ and\ \citenamefont {Singh}}]{Boyanovsky}%
  \BibitemOpen
  \bibfield  {author} {\bibinfo {author} {\bibfnamefont {D.}~\bibnamefont
  {Boyanovsky}}, \bibinfo {author} {\bibfnamefont {R.}~\bibnamefont {Holman}},
  \bibinfo {author} {\bibfnamefont {D.-S.}\ \bibnamefont {Lee}}, \bibinfo
  {author} {\bibfnamefont {J.~P.}\ \bibnamefont {Silva}}, \ and\ \bibinfo
  {author} {\bibfnamefont {A.}~\bibnamefont {Singh}},\ }\href {\doibase
  10.1016/0550-3213(95)00047-V} {\bibfield  {journal} {\bibinfo  {journal}
  {Nucl. Phys.}\ }\textbf {\bibinfo {volume} {B441}},\ \bibinfo {pages} {595}
  (\bibinfo {year} {1995})},\ \Eprint {http://arxiv.org/abs/hep-ph/9403267}
  {hep-ph/9403267} \BibitemShut {NoStop}%
\bibitem [{\citenamefont {Berry}\ and\ \citenamefont {Keating}(1999)}]{Berry}%
  \BibitemOpen
  \bibfield  {author} {\bibinfo {author} {\bibfnamefont {M.~V.}\ \bibnamefont
  {Berry}}\ and\ \bibinfo {author} {\bibfnamefont {J.~P.}\ \bibnamefont
  {Keating}},\ }\href@noop {} {\emph {\bibinfo {title} {Supersymmetry and Trace
  Formulae: Chaos and Disorder}}},\ edited by\ \bibinfo {editor} {\bibfnamefont
  {I.~V.}\ \bibnamefont {Lerner}}, \bibinfo {editor} {\bibfnamefont {J.~P.}\
  \bibnamefont {Keating}}, \ and\ \bibinfo {editor} {\bibfnamefont {D.~E.}\
  \bibnamefont {Khmelnitskii}}\ (\bibinfo  {publisher} {Springer US},\ \bibinfo
  {address} {Boston, MA},\ \bibinfo {year} {1999})\ pp.\ \bibinfo {pages}
  {355--367}\BibitemShut {NoStop}%
\bibitem [{\citenamefont {Sierra}\ and\ \citenamefont
  {Townsend}(2008)}]{Sierra}%
  \BibitemOpen
  \bibfield  {author} {\bibinfo {author} {\bibfnamefont {G.}~\bibnamefont
  {Sierra}}\ and\ \bibinfo {author} {\bibfnamefont {P.~K.}\ \bibnamefont
  {Townsend}},\ }\href {\doibase 10.1103/PhysRevLett.101.110201} {\bibfield
  {journal} {\bibinfo  {journal} {Phys. Rev. Lett.}\ }\textbf {\bibinfo
  {volume} {101}},\ \bibinfo {pages} {110201} (\bibinfo {year}
  {2008})}\BibitemShut {NoStop}%
\bibitem [{\citenamefont {Bhaduri}\ \emph {et~al.}(1995)\citenamefont
  {Bhaduri}, \citenamefont {Khare},\ and\ \citenamefont {Law}}]{Bhaduri}%
  \BibitemOpen
  \bibfield  {author} {\bibinfo {author} {\bibfnamefont {R.~K.}\ \bibnamefont
  {Bhaduri}}, \bibinfo {author} {\bibfnamefont {A.}~\bibnamefont {Khare}}, \
  and\ \bibinfo {author} {\bibfnamefont {J.}~\bibnamefont {Law}},\ }\href
  {\doibase 10.1103/PhysRevE.52.486} {\bibfield  {journal} {\bibinfo  {journal}
  {Phys. Rev. E}\ }\textbf {\bibinfo {volume} {52}},\ \bibinfo {pages} {486}
  (\bibinfo {year} {1995})}\BibitemShut {NoStop}%
\bibitem [{\citenamefont {Miller}\ and\ \citenamefont {Sarkar}(1998)}]{Miller}%
  \BibitemOpen
  \bibfield  {author} {\bibinfo {author} {\bibfnamefont {P.~A.}\ \bibnamefont
  {Miller}}\ and\ \bibinfo {author} {\bibfnamefont {S.}~\bibnamefont
  {Sarkar}},\ }\href {\doibase 10.1103/PhysRevE.58.4217} {\bibfield  {journal}
  {\bibinfo  {journal} {Phys. Rev. E}\ }\textbf {\bibinfo {volume} {58}},\
  \bibinfo {pages} {4217} (\bibinfo {year} {1998})}\BibitemShut {NoStop}%
\bibitem [{\citenamefont {Cucchietti}\ \emph {et~al.}(2003)\citenamefont
  {Cucchietti}, \citenamefont {Dalvit}, \citenamefont {Paz},\ and\
  \citenamefont {Zurek}}]{Cuchietti}%
  \BibitemOpen
  \bibfield  {author} {\bibinfo {author} {\bibfnamefont {F.~M.}\ \bibnamefont
  {Cucchietti}}, \bibinfo {author} {\bibfnamefont {D.~A.~R.}\ \bibnamefont
  {Dalvit}}, \bibinfo {author} {\bibfnamefont {J.~P.}\ \bibnamefont {Paz}}, \
  and\ \bibinfo {author} {\bibfnamefont {W.~H.}\ \bibnamefont {Zurek}},\ }\href
  {\doibase 10.1103/PhysRevLett.91.210403} {\bibfield  {journal} {\bibinfo
  {journal} {Phys. Rev. Lett.}\ }\textbf {\bibinfo {volume} {91}},\ \bibinfo
  {pages} {210403} (\bibinfo {year} {2003})}\BibitemShut {NoStop}%
\bibitem [{\citenamefont {Baskoutas}\ and\ \citenamefont
  {Jannussis}(1992)}]{Baskoutas}%
  \BibitemOpen
  \bibfield  {author} {\bibinfo {author} {\bibfnamefont {S.}~\bibnamefont
  {Baskoutas}}\ and\ \bibinfo {author} {\bibfnamefont {A.}~\bibnamefont
  {Jannussis}},\ }\href {http://stacks.iop.org/0305-4470/25/i=23/a=006}
  {\bibfield  {journal} {\bibinfo  {journal} {Journal of Physics A:
  Mathematical and General}\ }\textbf {\bibinfo {volume} {25}},\ \bibinfo
  {pages} {L1299} (\bibinfo {year} {1992})}\BibitemShut {NoStop}%
\bibitem [{\citenamefont {Gentilini}\ \emph {et~al.}(2015)\citenamefont
  {Gentilini}, \citenamefont {Braidotti}, \citenamefont {Marcucci},
  \citenamefont {DelRe},\ and\ \citenamefont {Conti}}]{Gentilini}%
  \BibitemOpen
  \bibfield  {author} {\bibinfo {author} {\bibfnamefont {S.}~\bibnamefont
  {Gentilini}}, \bibinfo {author} {\bibfnamefont {M.~C.}\ \bibnamefont
  {Braidotti}}, \bibinfo {author} {\bibfnamefont {G.}~\bibnamefont {Marcucci}},
  \bibinfo {author} {\bibfnamefont {E.}~\bibnamefont {DelRe}}, \ and\ \bibinfo
  {author} {\bibfnamefont {C.}~\bibnamefont {Conti}},\ }\href@noop {}
  {\bibfield  {journal} {\bibinfo  {journal} {Scientific Reports}\ }\textbf
  {\bibinfo {volume} {5}},\ \bibinfo {pages} {15816} (\bibinfo {year}
  {2015})}\BibitemShut {NoStop}%
\bibitem [{\citenamefont {{Friess}}\ and\ \citenamefont
  {{Verlinde}}(2004)}]{Friess}%
  \BibitemOpen
  \bibfield  {author} {\bibinfo {author} {\bibfnamefont {J.~J.}\ \bibnamefont
  {{Friess}}}\ and\ \bibinfo {author} {\bibfnamefont {H.}~\bibnamefont
  {{Verlinde}}},\ }\href@noop {} {\bibfield  {journal} {\bibinfo  {journal}
  {ArXiv High Energy Physics - Theory e-prints}\ } (\bibinfo {year} {2004})},\
  \Eprint {http://arxiv.org/abs/hep-th/0411100} {hep-th/0411100} \BibitemShut
  {NoStop}%
\bibitem [{\citenamefont {Maldacena}\ and\ \citenamefont
  {Seiberg}(2005)}]{Maldacena05}%
  \BibitemOpen
  \bibfield  {author} {\bibinfo {author} {\bibfnamefont {J.~M.}\ \bibnamefont
  {Maldacena}}\ and\ \bibinfo {author} {\bibfnamefont {N.}~\bibnamefont
  {Seiberg}},\ }\href@noop {} {\bibfield  {journal} {\bibinfo  {journal}
  {JHEP}\ }\textbf {\bibinfo {volume} {09}},\ \bibinfo {pages} {077} (\bibinfo
  {year} {2005})}\BibitemShut {NoStop}%
\bibitem [{\citenamefont {Cremonini}(2005)}]{Cremonini}%
  \BibitemOpen
  \bibfield  {author} {\bibinfo {author} {\bibfnamefont {S.}~\bibnamefont
  {Cremonini}},\ }\href {http://stacks.iop.org/1126-6708/2005/i=10/a=014}
  {\bibfield  {journal} {\bibinfo  {journal} {Journal of High Energy Physics}\
  }\textbf {\bibinfo {volume} {2005}},\ \bibinfo {pages} {014} (\bibinfo {year}
  {2005})}\BibitemShut {NoStop}%
\bibitem [{\citenamefont {Schoutens}\ \emph {et~al.}(1993)\citenamefont
  {Schoutens}, \citenamefont {Verlinde},\ and\ \citenamefont
  {Verlinde}}]{Schoutens}%
  \BibitemOpen
  \bibfield  {author} {\bibinfo {author} {\bibfnamefont {K.}~\bibnamefont
  {Schoutens}}, \bibinfo {author} {\bibfnamefont {H.}~\bibnamefont {Verlinde}},
  \ and\ \bibinfo {author} {\bibfnamefont {E.}~\bibnamefont {Verlinde}},\
  }\href {\doibase 10.1103/PhysRevD.48.2670} {\bibfield  {journal} {\bibinfo
  {journal} {Phys. Rev. D}\ }\textbf {\bibinfo {volume} {48}},\ \bibinfo
  {pages} {2670} (\bibinfo {year} {1993})}\BibitemShut {NoStop}%
\bibitem [{\citenamefont {Guth}\ and\ \citenamefont {Pi}(1985)}]{Guth}%
  \BibitemOpen
  \bibfield  {author} {\bibinfo {author} {\bibfnamefont {A.~H.}\ \bibnamefont
  {Guth}}\ and\ \bibinfo {author} {\bibfnamefont {S.-Y.}\ \bibnamefont {Pi}},\
  }\href {\doibase 10.1103/PhysRevD.32.1899} {\bibfield  {journal} {\bibinfo
  {journal} {Phys. Rev. D}\ }\textbf {\bibinfo {volume} {32}},\ \bibinfo
  {pages} {1899} (\bibinfo {year} {1985})}\BibitemShut {NoStop}%
\bibitem [{\citenamefont {Sewell}(1982)}]{Sewell82}%
  \BibitemOpen
  \bibfield  {author} {\bibinfo {author} {\bibfnamefont {G.~L.}\ \bibnamefont
  {Sewell}},\ }\href {\doibase https://doi.org/10.1016/0003-4916(82)90285-8}
  {\bibfield  {journal} {\bibinfo  {journal} {Annals of Physics}\ }\textbf
  {\bibinfo {volume} {141}},\ \bibinfo {pages} {201 } (\bibinfo {year}
  {1982})}\BibitemShut {NoStop}%
\bibitem [{\citenamefont {{Stone}}(2013)}]{Stone12}%
  \BibitemOpen
  \bibfield  {author} {\bibinfo {author} {\bibfnamefont {M.}~\bibnamefont
  {{Stone}}},\ }\href@noop {} {\bibfield  {journal} {\bibinfo  {journal}
  {Classical and Quantum Gravity}\ }\textbf {\bibinfo {volume} {30}} (\bibinfo
  {year} {2013})},\ \Eprint {http://arxiv.org/abs/1209.2317} {arXiv:1209.2317
  [gr-qc]} \BibitemShut {NoStop}%
\bibitem [{\citenamefont {{Chandrasekhar}}(1983)}]{Chandrasekhar}%
  \BibitemOpen
  \bibfield  {author} {\bibinfo {author} {\bibfnamefont {S.}~\bibnamefont
  {{Chandrasekhar}}},\ }\href@noop {} {\emph {\bibinfo {title} {{The
  mathematical theory of black holes}}}}\ (\bibinfo  {publisher} {Clarendon
  Press/Oxford University Press (International Series of Monographs on
  Physics.~Volume 69)},\ \bibinfo {year} {1983})\BibitemShut {NoStop}%
\bibitem [{\citenamefont {Konoplya}\ and\ \citenamefont
  {Zhidenko}(2011)}]{Konoplaya11}%
  \BibitemOpen
  \bibfield  {author} {\bibinfo {author} {\bibfnamefont {R.~A.}\ \bibnamefont
  {Konoplya}}\ and\ \bibinfo {author} {\bibfnamefont {A.}~\bibnamefont
  {Zhidenko}},\ }\href@noop {} {\bibfield  {journal} {\bibinfo  {journal} {Rev.
  Mod. Phys.}\ }\textbf {\bibinfo {volume} {83}},\ \bibinfo {pages} {793}
  (\bibinfo {year} {2011})}\BibitemShut {NoStop}%
\bibitem [{\citenamefont {Misner}\ \emph {et~al.}(1973)\citenamefont {Misner},
  \citenamefont {Thorne},\ and\ \citenamefont {Wheeler}}]{MTW}%
  \BibitemOpen
  \bibfield  {author} {\bibinfo {author} {\bibfnamefont {C.~W.}\ \bibnamefont
  {Misner}}, \bibinfo {author} {\bibfnamefont {K.~S.}\ \bibnamefont {Thorne}},
  \ and\ \bibinfo {author} {\bibfnamefont {J.~A.}\ \bibnamefont {Wheeler}},\
  }\href@noop {} {\emph {\bibinfo {title} {Gravitation}}}\ (\bibinfo
  {publisher} {New York : W.H. Freeman and Company},\ \bibinfo {year}
  {1973})\BibitemShut {NoStop}%
\bibitem [{\citenamefont {Regge}\ and\ \citenamefont
  {Wheeler}(1957)}]{ReggeWheeler}%
  \BibitemOpen
  \bibfield  {author} {\bibinfo {author} {\bibfnamefont {T.}~\bibnamefont
  {Regge}}\ and\ \bibinfo {author} {\bibfnamefont {J.~A.}\ \bibnamefont
  {Wheeler}},\ }\href {\doibase 10.1103/PhysRev.108.1063} {\bibfield  {journal}
  {\bibinfo  {journal} {Phys. Rev.}\ }\textbf {\bibinfo {volume} {108}},\
  \bibinfo {pages} {1063} (\bibinfo {year} {1957})}\BibitemShut {NoStop}%
\bibitem [{\citenamefont {Andersson}\ and\ \citenamefont
  {Jensen}(2001)}]{Andersson}%
  \BibitemOpen
  \bibfield  {author} {\bibinfo {author} {\bibfnamefont {N.}~\bibnamefont
  {Andersson}}\ and\ \bibinfo {author} {\bibfnamefont {B.}~\bibnamefont
  {Jensen}},\ }\href {https://eprints.soton.ac.uk/29436/} {\emph {\bibinfo
  {title} {Scattering, Two Volume Set: Scattering and Inverse Scattering in
  Pure and Applied Science}}},\ edited by\ \bibinfo {editor} {\bibfnamefont
  {E.}~\bibnamefont {Pike}}\ and\ \bibinfo {editor} {\bibfnamefont
  {P.}~\bibnamefont {Sabatier}}\ (\bibinfo  {publisher} {Academic Press},\
  \bibinfo {year} {2001})\BibitemShut {NoStop}%
\bibitem [{\citenamefont {Vishveshwara}(1970)}]{Vishveshwara}%
  \BibitemOpen
  \bibfield  {author} {\bibinfo {author} {\bibfnamefont {C.~V.}\ \bibnamefont
  {Vishveshwara}},\ }\href@noop {} {\bibfield  {journal} {\bibinfo  {journal}
  {Nature}\ }\textbf {\bibinfo {volume} {227}},\ \bibinfo {pages} {936 EP}
  (\bibinfo {year} {1970})}\BibitemShut {NoStop}%
\bibitem [{\citenamefont {{Schutz}}\ and\ \citenamefont
  {{Will}}(1985)}]{Schutz85}%
  \BibitemOpen
  \bibfield  {author} {\bibinfo {author} {\bibfnamefont {B.~F.}\ \bibnamefont
  {{Schutz}}}\ and\ \bibinfo {author} {\bibfnamefont {C.~M.}\ \bibnamefont
  {{Will}}},\ }\href@noop {} {\bibfield  {journal} {\bibinfo  {journal} {apjl}\
  }\textbf {\bibinfo {volume} {291}},\ \bibinfo {pages} {L33} (\bibinfo {year}
  {1985})}\BibitemShut {NoStop}%
\bibitem [{\citenamefont {Ford}\ and\ \citenamefont {Wheeler}(1959)}]{Ford59}%
  \BibitemOpen
  \bibfield  {author} {\bibinfo {author} {\bibfnamefont {K.~W.}\ \bibnamefont
  {Ford}}\ and\ \bibinfo {author} {\bibfnamefont {J.~A.}\ \bibnamefont
  {Wheeler}},\ }\href@noop {} {\bibfield  {journal} {\bibinfo  {journal}
  {Annals of Physics}\ }\textbf {\bibinfo {volume} {7}},\ \bibinfo {pages} {259
  } (\bibinfo {year} {1959})}\BibitemShut {NoStop}%
\bibitem [{\citenamefont {Liu}\ and\ \citenamefont
  {Mashhoon}(1996)}]{Mashhoon}%
  \BibitemOpen
  \bibfield  {author} {\bibinfo {author} {\bibfnamefont {H.}~\bibnamefont
  {Liu}}\ and\ \bibinfo {author} {\bibfnamefont {B.}~\bibnamefont {Mashhoon}},\
  }\href@noop {} {\bibfield  {journal} {\bibinfo  {journal} {Classical and
  Quantum Gravity}\ }\textbf {\bibinfo {volume} {13}},\ \bibinfo {pages} {233}
  (\bibinfo {year} {1996})}\BibitemShut {NoStop}%
\bibitem [{\citenamefont {{Betzios}}\ \emph {et~al.}(2016)\citenamefont
  {{Betzios}}, \citenamefont {{Gaddam}},\ and\ \citenamefont
  {{Papadoulaki}}}]{Betzios16}%
  \BibitemOpen
  \bibfield  {author} {\bibinfo {author} {\bibfnamefont {P.}~\bibnamefont
  {{Betzios}}}, \bibinfo {author} {\bibfnamefont {N.}~\bibnamefont {{Gaddam}}},
  \ and\ \bibinfo {author} {\bibfnamefont {O.}~\bibnamefont {{Papadoulaki}}},\
  }\href@noop {} {\bibfield  {journal} {\bibinfo  {journal} {Journal of High
  Energy Physics}\ }\textbf {\bibinfo {volume} {11}},\ \bibinfo {pages} {131}
  (\bibinfo {year} {2016})},\ \Eprint {http://arxiv.org/abs/1607.07885}
  {arXiv:1607.07885 [hep-th]} \BibitemShut {NoStop}%
\bibitem [{\citenamefont {Dray}\ and\ \citenamefont {'t~Hooft}(1985)}]{Dray85}%
  \BibitemOpen
  \bibfield  {author} {\bibinfo {author} {\bibfnamefont {T.}~\bibnamefont
  {Dray}}\ and\ \bibinfo {author} {\bibfnamefont {G.}~\bibnamefont
  {'t~Hooft}},\ }\href@noop {} {\bibfield  {journal} {\bibinfo  {journal}
  {Comm. Math. Phys.}\ }\textbf {\bibinfo {volume} {99}},\ \bibinfo {pages}
  {613} (\bibinfo {year} {1985})}\BibitemShut {NoStop}%
\bibitem [{\citenamefont {Fertig}\ and\ \citenamefont
  {Halperin}(1987)}]{Fertig87}%
  \BibitemOpen
  \bibfield  {author} {\bibinfo {author} {\bibfnamefont {H.~A.}\ \bibnamefont
  {Fertig}}\ and\ \bibinfo {author} {\bibfnamefont {B.~I.}\ \bibnamefont
  {Halperin}},\ }\href@noop {} {\bibfield  {journal} {\bibinfo  {journal}
  {Phys. Rev. B}\ }\textbf {\bibinfo {volume} {36}},\ \bibinfo {pages} {7969}
  (\bibinfo {year} {1987})}\BibitemShut {NoStop}%
\bibitem [{\citenamefont {B\"uttiker}(1990)}]{Buttiker90}%
  \BibitemOpen
  \bibfield  {author} {\bibinfo {author} {\bibfnamefont {M.}~\bibnamefont
  {B\"uttiker}},\ }\href@noop {} {\bibfield  {journal} {\bibinfo  {journal}
  {Phys. Rev. B}\ }\textbf {\bibinfo {volume} {41}},\ \bibinfo {pages} {7906}
  (\bibinfo {year} {1990})}\BibitemShut {NoStop}%
\bibitem [{\citenamefont {Matthews}\ and\ \citenamefont
  {Cooper}(2009)}]{Matthews}%
  \BibitemOpen
  \bibfield  {author} {\bibinfo {author} {\bibfnamefont {A.}~\bibnamefont
  {Matthews}}\ and\ \bibinfo {author} {\bibfnamefont {N.~R.}\ \bibnamefont
  {Cooper}},\ }\href {\doibase 10.1103/PhysRevB.80.165309} {\bibfield
  {journal} {\bibinfo  {journal} {Phys. Rev. B}\ }\textbf {\bibinfo {volume}
  {80}},\ \bibinfo {pages} {165309} (\bibinfo {year} {2009})}\BibitemShut
  {NoStop}%
\bibitem [{\citenamefont {Vishveshwara}\ and\ \citenamefont
  {Cooper}(2010)}]{Smitha10}%
  \BibitemOpen
  \bibfield  {author} {\bibinfo {author} {\bibfnamefont {S.}~\bibnamefont
  {Vishveshwara}}\ and\ \bibinfo {author} {\bibfnamefont {N.~R.}\ \bibnamefont
  {Cooper}},\ }\href@noop {} {\bibfield  {journal} {\bibinfo  {journal} {Phys.
  Rev. B}\ }\textbf {\bibinfo {volume} {81}},\ \bibinfo {pages} {201306}
  (\bibinfo {year} {2010})}\BibitemShut {NoStop}%
\bibitem [{\citenamefont {Girvin}\ and\ \citenamefont
  {Jach}(1984)}]{girvinjach}%
  \BibitemOpen
  \bibfield  {author} {\bibinfo {author} {\bibfnamefont {S.~M.}\ \bibnamefont
  {Girvin}}\ and\ \bibinfo {author} {\bibfnamefont {T.}~\bibnamefont {Jach}},\
  }\href {\doibase 10.1103/PhysRevB.29.5617} {\bibfield  {journal} {\bibinfo
  {journal} {Phys. Rev. B}\ }\textbf {\bibinfo {volume} {29}},\ \bibinfo
  {pages} {5617} (\bibinfo {year} {1984})}\BibitemShut {NoStop}%
\bibitem [{\citenamefont {Read}\ and\ \citenamefont {Rezayi}(2011)}]{ReadVisc}%
  \BibitemOpen
  \bibfield  {author} {\bibinfo {author} {\bibfnamefont {N.}~\bibnamefont
  {Read}}\ and\ \bibinfo {author} {\bibfnamefont {E.~H.}\ \bibnamefont
  {Rezayi}},\ }\href@noop {} {\bibfield  {journal} {\bibinfo  {journal} {Phys.
  Rev. B}\ }\textbf {\bibinfo {volume} {84}},\ \bibinfo {pages} {085316}
  (\bibinfo {year} {2011})}\BibitemShut {NoStop}%
\bibitem [{\citenamefont {Bradlyn}\ \emph {et~al.}(2012)\citenamefont
  {Bradlyn}, \citenamefont {Goldstein},\ and\ \citenamefont
  {Read}}]{Bradlyn2012}%
  \BibitemOpen
  \bibfield  {author} {\bibinfo {author} {\bibfnamefont {B.}~\bibnamefont
  {Bradlyn}}, \bibinfo {author} {\bibfnamefont {M.}~\bibnamefont {Goldstein}},
  \ and\ \bibinfo {author} {\bibfnamefont {N.}~\bibnamefont {Read}},\ }\href
  {\doibase 10.1103/PhysRevB.86.245309} {\bibfield  {journal} {\bibinfo
  {journal} {Phys. Rev. B}\ }\textbf {\bibinfo {volume} {86}},\ \bibinfo
  {pages} {245309} (\bibinfo {year} {2012})}\BibitemShut {NoStop}%
\bibitem [{\citenamefont {Yuen}(1976)}]{yuensqueezing}%
  \BibitemOpen
  \bibfield  {author} {\bibinfo {author} {\bibfnamefont {H.~P.}\ \bibnamefont
  {Yuen}},\ }\href {\doibase 10.1103/PhysRevA.13.2226} {\bibfield  {journal}
  {\bibinfo  {journal} {Phys. Rev. A}\ }\textbf {\bibinfo {volume} {13}},\
  \bibinfo {pages} {2226} (\bibinfo {year} {1976})}\BibitemShut {NoStop}%
\bibitem [{\citenamefont {Francesco}\ \emph {et~al.}(1997)\citenamefont
  {Francesco}, \citenamefont {Mathieu},\ and\ \citenamefont
  {Senechal}}]{FrancescoCFT}%
  \BibitemOpen
  \bibfield  {author} {\bibinfo {author} {\bibfnamefont {P.}~\bibnamefont
  {Francesco}}, \bibinfo {author} {\bibfnamefont {P.}~\bibnamefont {Mathieu}},
  \ and\ \bibinfo {author} {\bibfnamefont {D.}~\bibnamefont {Senechal}},\
  }\href@noop {} {\emph {\bibinfo {title} {Conformal Field Theory}}}\ (\bibinfo
   {publisher} {Springer-Verlag New York},\ \bibinfo {year} {1997})\BibitemShut
  {NoStop}%
\bibitem [{\citenamefont {Moses}\ and\ \citenamefont
  {Quesada}(1972)}]{Moses72}%
  \BibitemOpen
  \bibfield  {author} {\bibinfo {author} {\bibfnamefont {H.~E.}\ \bibnamefont
  {Moses}}\ and\ \bibinfo {author} {\bibfnamefont {A.~F.}\ \bibnamefont
  {Quesada}},\ }\href@noop {} {\bibfield  {journal} {\bibinfo  {journal}
  {Archive for Rational Mechanics and Analysis}\ }\textbf {\bibinfo {volume}
  {44}},\ \bibinfo {pages} {217} (\bibinfo {year} {1972})}\BibitemShut
  {NoStop}%
\bibitem [{\citenamefont {Cycon}\ \emph {et~al.}(1987)\citenamefont {Cycon},
  \citenamefont {Froese}, \citenamefont {Kirsch},\ and\ \citenamefont
  {Simon}}]{Cycon87}%
  \BibitemOpen
  \bibfield  {author} {\bibinfo {author} {\bibfnamefont {H.}~\bibnamefont
  {Cycon}}, \bibinfo {author} {\bibfnamefont {R.~G.}\ \bibnamefont {Froese}},
  \bibinfo {author} {\bibfnamefont {W.}~\bibnamefont {Kirsch}}, \ and\ \bibinfo
  {author} {\bibfnamefont {B.}~\bibnamefont {Simon}},\ }\href@noop {} {\emph
  {\bibinfo {title} {Schrödinger Operators (With Applications to Quantum
  Mechanics and Global Geometry)}}}\ (\bibinfo  {publisher} {Springer-Verlag
  Berlin Heidelberg},\ \bibinfo {year} {1987})\BibitemShut {NoStop}%
\bibitem [{\citenamefont {Chalker}\ and\ \citenamefont
  {Coddington}(1988)}]{Chalker}%
  \BibitemOpen
  \bibfield  {author} {\bibinfo {author} {\bibfnamefont {J.~T.}\ \bibnamefont
  {Chalker}}\ and\ \bibinfo {author} {\bibfnamefont {P.~D.}\ \bibnamefont
  {Coddington}},\ }\href@noop {} {\bibfield  {journal} {\bibinfo  {journal} {J.
  Phys. C: Solid State Phys}\ }\textbf {\bibinfo {volume} {21}} (\bibinfo
  {year} {1988})}\BibitemShut {NoStop}%
\bibitem [{\citenamefont {Huckestein}(1995)}]{Huckestein}%
  \BibitemOpen
  \bibfield  {author} {\bibinfo {author} {\bibfnamefont {B.}~\bibnamefont
  {Huckestein}},\ }\href {\doibase 10.1103/RevModPhys.67.357} {\bibfield
  {journal} {\bibinfo  {journal} {Rev. Mod. Phys.}\ }\textbf {\bibinfo {volume}
  {67}},\ \bibinfo {pages} {357} (\bibinfo {year} {1995})}\BibitemShut
  {NoStop}%
\bibitem [{\citenamefont {Sen}\ \emph {et~al.}(2008)\citenamefont {Sen},
  \citenamefont {Stone},\ and\ \citenamefont {Vishveshwara}}]{Sen}%
  \BibitemOpen
  \bibfield  {author} {\bibinfo {author} {\bibfnamefont {D.}~\bibnamefont
  {Sen}}, \bibinfo {author} {\bibfnamefont {M.}~\bibnamefont {Stone}}, \ and\
  \bibinfo {author} {\bibfnamefont {S.}~\bibnamefont {Vishveshwara}},\ }\href
  {\doibase 10.1103/PhysRevB.77.115442} {\bibfield  {journal} {\bibinfo
  {journal} {Phys. Rev. B}\ }\textbf {\bibinfo {volume} {77}},\ \bibinfo
  {pages} {115442} (\bibinfo {year} {2008})}\BibitemShut {NoStop}%
\bibitem [{\citenamefont {Parikh}\ and\ \citenamefont
  {Wilczek}(2000)}]{Parikh00}%
  \BibitemOpen
  \bibfield  {author} {\bibinfo {author} {\bibfnamefont {M.~K.}\ \bibnamefont
  {Parikh}}\ and\ \bibinfo {author} {\bibfnamefont {F.}~\bibnamefont
  {Wilczek}},\ }\href@noop {} {\bibfield  {journal} {\bibinfo  {journal} {Phys.
  Rev. Lett.}\ }\textbf {\bibinfo {volume} {85}},\ \bibinfo {pages} {5042}
  (\bibinfo {year} {2000})}\BibitemShut {NoStop}%
\bibitem [{\citenamefont {{Vanzo}}\ \emph {et~al.}(2011)\citenamefont
  {{Vanzo}}, \citenamefont {{Acquaviva}},\ and\ \citenamefont {{Di
  Criscienzo}}}]{Vanzo11}%
  \BibitemOpen
  \bibfield  {author} {\bibinfo {author} {\bibfnamefont {L.}~\bibnamefont
  {{Vanzo}}}, \bibinfo {author} {\bibfnamefont {G.}~\bibnamefont
  {{Acquaviva}}}, \ and\ \bibinfo {author} {\bibfnamefont {R.}~\bibnamefont
  {{Di Criscienzo}}},\ }\href@noop {} {\bibfield  {journal} {\bibinfo
  {journal} {Classical and Quantum Gravity}\ }\textbf {\bibinfo {volume}
  {28}},\ \bibinfo {eid} {183001} (\bibinfo {year} {2011})}\BibitemShut
  {NoStop}%
\bibitem [{\citenamefont {Bisognano}\ and\ \citenamefont
  {Wichmann}(1976)}]{Bisognano}%
  \BibitemOpen
  \bibfield  {author} {\bibinfo {author} {\bibfnamefont {J.~J.}\ \bibnamefont
  {Bisognano}}\ and\ \bibinfo {author} {\bibfnamefont {E.~H.}\ \bibnamefont
  {Wichmann}},\ }\href {\doibase 10.1063/1.522898} {\bibfield  {journal}
  {\bibinfo  {journal} {Journal of Mathematical Physics}\ }\textbf {\bibinfo
  {volume} {17}},\ \bibinfo {pages} {303} (\bibinfo {year} {1976})},\ \Eprint
  {http://arxiv.org/abs/https://aip.scitation.org/doi/pdf/10.1063/1.522898}
  {https://aip.scitation.org/doi/pdf/10.1063/1.522898} \BibitemShut {NoStop}%
\bibitem [{\citenamefont {Robinson}\ and\ \citenamefont
  {Wilczek}(2005)}]{Robinson05}%
  \BibitemOpen
  \bibfield  {author} {\bibinfo {author} {\bibfnamefont {S.~P.}\ \bibnamefont
  {Robinson}}\ and\ \bibinfo {author} {\bibfnamefont {F.}~\bibnamefont
  {Wilczek}},\ }\href {\doibase 10.1103/PhysRevLett.95.011303} {\bibfield
  {journal} {\bibinfo  {journal} {Phys. Rev. Lett.}\ }\textbf {\bibinfo
  {volume} {95}},\ \bibinfo {pages} {011303} (\bibinfo {year}
  {2005})}\BibitemShut {NoStop}%
\bibitem [{\citenamefont {Hawking}(1974)}]{Hawking}%
  \BibitemOpen
  \bibfield  {author} {\bibinfo {author} {\bibfnamefont {S.~W.}\ \bibnamefont
  {Hawking}},\ }\href@noop {} {\bibfield  {journal} {\bibinfo  {journal}
  {Nature}\ }\textbf {\bibinfo {volume} {248}},\ \bibinfo {pages} {30 EP}
  (\bibinfo {year} {1974})}\BibitemShut {NoStop}%
\bibitem [{\citenamefont {Hawking}(1975)}]{Hawking75}%
  \BibitemOpen
  \bibfield  {author} {\bibinfo {author} {\bibfnamefont {S.~W.}\ \bibnamefont
  {Hawking}},\ }\href@noop {} {\bibfield  {journal} {\bibinfo  {journal}
  {Communications in Mathematical Physics}\ }\textbf {\bibinfo {volume} {43}},\
  \bibinfo {pages} {199} (\bibinfo {year} {1975})}\BibitemShut {NoStop}%
\bibitem [{\citenamefont {Unruh}(1976)}]{Unruh76}%
  \BibitemOpen
  \bibfield  {author} {\bibinfo {author} {\bibfnamefont {W.~G.}\ \bibnamefont
  {Unruh}},\ }\href@noop {} {\bibfield  {journal} {\bibinfo  {journal} {Phys.
  Rev. D}\ }\textbf {\bibinfo {volume} {14}},\ \bibinfo {pages} {870} (\bibinfo
  {year} {1976})}\BibitemShut {NoStop}%
\bibitem [{\citenamefont {Volovik}(1999)}]{volovik1999vierbein}%
  \BibitemOpen
  \bibfield  {author} {\bibinfo {author} {\bibfnamefont {G.}~\bibnamefont
  {Volovik}},\ }\href@noop {} {\bibfield  {journal} {\bibinfo  {journal}
  {Journal of Experimental and Theoretical Physics Letters}\ }\textbf {\bibinfo
  {volume} {70}},\ \bibinfo {pages} {711} (\bibinfo {year} {1999})}\BibitemShut
  {NoStop}%
\bibitem [{\citenamefont {Mukhanov}\ and\ \citenamefont
  {Winitzki}(2007)}]{Mukhanov}%
  \BibitemOpen
  \bibfield  {author} {\bibinfo {author} {\bibfnamefont {V.}~\bibnamefont
  {Mukhanov}}\ and\ \bibinfo {author} {\bibfnamefont {S.}~\bibnamefont
  {Winitzki}},\ }\href@noop {} {\emph {\bibinfo {title} {Quantum effects in
  gravity}}}\ (\bibinfo  {publisher} {Cambridge university press},\ \bibinfo
  {year} {2007})\BibitemShut {NoStop}%
\bibitem [{\citenamefont {Crispino}\ \emph {et~al.}(2008)\citenamefont
  {Crispino}, \citenamefont {Higuchi},\ and\ \citenamefont
  {Matsas}}]{Crispino08}%
  \BibitemOpen
  \bibfield  {author} {\bibinfo {author} {\bibfnamefont {L.~C.~B.}\
  \bibnamefont {Crispino}}, \bibinfo {author} {\bibfnamefont {A.}~\bibnamefont
  {Higuchi}}, \ and\ \bibinfo {author} {\bibfnamefont {G.~E.~A.}\ \bibnamefont
  {Matsas}},\ }\href@noop {} {\bibfield  {journal} {\bibinfo  {journal} {Rev.
  Mod. Phys.}\ }\textbf {\bibinfo {volume} {80}},\ \bibinfo {pages} {787}
  (\bibinfo {year} {2008})}\BibitemShut {NoStop}%
\bibitem [{\citenamefont {Alsing}\ and\ \citenamefont
  {Milonni}(2004)}]{Alsing}%
  \BibitemOpen
  \bibfield  {author} {\bibinfo {author} {\bibfnamefont {P.~M.}\ \bibnamefont
  {Alsing}}\ and\ \bibinfo {author} {\bibfnamefont {P.~W.}\ \bibnamefont
  {Milonni}},\ }\href {\doibase 10.1119/1.1761064} {\bibfield  {journal}
  {\bibinfo  {journal} {American Journal of Physics}\ }\textbf {\bibinfo
  {volume} {72}},\ \bibinfo {pages} {1524} (\bibinfo {year} {2004})},\ \Eprint
  {http://arxiv.org/abs/https://doi.org/10.1119/1.1761064}
  {https://doi.org/10.1119/1.1761064} \BibitemShut {NoStop}%
\bibitem [{\citenamefont {Arzano}\ and\ \citenamefont
  {Kowalski-Glikman}(2019)}]{Arzano}%
  \BibitemOpen
  \bibfield  {author} {\bibinfo {author} {\bibfnamefont {M.}~\bibnamefont
  {Arzano}}\ and\ \bibinfo {author} {\bibfnamefont {J.}~\bibnamefont
  {Kowalski-Glikman}},\ }\href {\doibase
  https://doi.org/10.1016/j.physletb.2018.10.019} {\bibfield  {journal}
  {\bibinfo  {journal} {Physics Letters B}\ }\textbf {\bibinfo {volume}
  {788}},\ \bibinfo {pages} {82 } (\bibinfo {year} {2019})}\BibitemShut
  {NoStop}%
\bibitem [{\citenamefont {Press}(1971)}]{Press}%
  \BibitemOpen
  \bibfield  {author} {\bibinfo {author} {\bibfnamefont {W.~H.}\ \bibnamefont
  {Press}},\ }\href {\doibase 10.1086/180849} {\bibfield  {journal} {\bibinfo
  {journal} {Astrophys. J.}\ }\textbf {\bibinfo {volume} {170}},\ \bibinfo
  {pages} {L105} (\bibinfo {year} {1971})}\BibitemShut {NoStop}%
\bibitem [{\citenamefont {Kokkotas}\ and\ \citenamefont
  {Schmidt}(1999)}]{Kokkotas}%
  \BibitemOpen
  \bibfield  {author} {\bibinfo {author} {\bibfnamefont {K.~D.}\ \bibnamefont
  {Kokkotas}}\ and\ \bibinfo {author} {\bibfnamefont {B.~G.}\ \bibnamefont
  {Schmidt}},\ }\href@noop {} {\bibfield  {journal} {\bibinfo  {journal}
  {Living Reviews in Relativity}\ }\textbf {\bibinfo {volume} {2}},\ \bibinfo
  {pages} {2} (\bibinfo {year} {1999})}\BibitemShut {NoStop}%
\bibitem [{\citenamefont {Berti}\ \emph {et~al.}(2009)\citenamefont {Berti},
  \citenamefont {Cardoso},\ and\ \citenamefont {Starinets}}]{Berti}%
  \BibitemOpen
  \bibfield  {author} {\bibinfo {author} {\bibfnamefont {E.}~\bibnamefont
  {Berti}}, \bibinfo {author} {\bibfnamefont {V.}~\bibnamefont {Cardoso}}, \
  and\ \bibinfo {author} {\bibfnamefont {A.~O.}\ \bibnamefont {Starinets}},\
  }\href {http://stacks.iop.org/0264-9381/26/i=16/a=163001} {\bibfield
  {journal} {\bibinfo  {journal} {Classical and Quantum Gravity}\ }\textbf
  {\bibinfo {volume} {26}},\ \bibinfo {pages} {163001} (\bibinfo {year}
  {2009})}\BibitemShut {NoStop}%
\bibitem [{\citenamefont {Horowitz}\ and\ \citenamefont
  {Hubeny}(2000)}]{Hubeny}%
  \BibitemOpen
  \bibfield  {author} {\bibinfo {author} {\bibfnamefont {G.~T.}\ \bibnamefont
  {Horowitz}}\ and\ \bibinfo {author} {\bibfnamefont {V.~E.}\ \bibnamefont
  {Hubeny}},\ }\href {\doibase 10.1103/PhysRevD.62.024027} {\bibfield
  {journal} {\bibinfo  {journal} {Phys. Rev. D}\ }\textbf {\bibinfo {volume}
  {62}},\ \bibinfo {pages} {024027} (\bibinfo {year} {2000})}\BibitemShut
  {NoStop}%
\bibitem [{\citenamefont {Panotopoulos}\ and\ \citenamefont {Ãngel
  RincÃ³n}(2017)}]{Panotopoulos}%
  \BibitemOpen
  \bibfield  {author} {\bibinfo {author} {\bibfnamefont {G.}~\bibnamefont
  {Panotopoulos}}\ and\ \bibinfo {author} {\bibnamefont {Ãngel RincÃ³n}},\
  }\href {\doibase https://doi.org/10.1016/j.physletb.2017.07.014} {\bibfield
  {journal} {\bibinfo  {journal} {Physics Letters B}\ }\textbf {\bibinfo
  {volume} {772}},\ \bibinfo {pages} {523 } (\bibinfo {year}
  {2017})}\BibitemShut {NoStop}%
\bibitem [{\citenamefont {{Teukolsky}}(1973)}]{Teukolsky}%
  \BibitemOpen
  \bibfield  {author} {\bibinfo {author} {\bibfnamefont {S.~A.}\ \bibnamefont
  {{Teukolsky}}},\ }\href {\doibase 10.1086/152444} {\bibfield  {journal}
  {\bibinfo  {journal} {\apj}\ }\textbf {\bibinfo {volume} {185}},\ \bibinfo
  {pages} {635} (\bibinfo {year} {1973})}\BibitemShut {NoStop}%
\bibitem [{\citenamefont {Chandrasekhar}\ and\ \citenamefont
  {Detweiler}(1975)}]{ChandraDetweiler}%
  \BibitemOpen
  \bibfield  {author} {\bibinfo {author} {\bibfnamefont {S.}~\bibnamefont
  {Chandrasekhar}}\ and\ \bibinfo {author} {\bibfnamefont {S.~L.}\ \bibnamefont
  {Detweiler}},\ }\href {\doibase 10.1098/rspa.1975.0112} {\bibfield  {journal}
  {\bibinfo  {journal} {Proc. Roy. Soc. Lond.}\ }\textbf {\bibinfo {volume}
  {A344}},\ \bibinfo {pages} {441} (\bibinfo {year} {1975})}\BibitemShut
  {NoStop}%
\bibitem [{\citenamefont {Perelomov}\ and\ \citenamefont
  {Zelâdovich}(1998)}]{Perelomov}%
  \BibitemOpen
  \bibfield  {author} {\bibinfo {author} {\bibfnamefont {A.~M.}\ \bibnamefont
  {Perelomov}}\ and\ \bibinfo {author} {\bibfnamefont {Y.~B.}\ \bibnamefont
  {Zelâdovich}},\ }\href@noop {} {\emph {\bibinfo {title} {Quantum
  Mechanics}}}\ (\bibinfo  {publisher} {WORLD SCIENTIFIC},\ \bibinfo {year}
  {1998})\BibitemShut {NoStop}%
\bibitem [{\citenamefont {Parravicini}\ \emph {et~al.}(1980)\citenamefont
  {Parravicini}, \citenamefont {Gorini},\ and\ \citenamefont
  {Sudarshan}}]{Parravacini}%
  \BibitemOpen
  \bibfield  {author} {\bibinfo {author} {\bibfnamefont {G.}~\bibnamefont
  {Parravicini}}, \bibinfo {author} {\bibfnamefont {V.}~\bibnamefont {Gorini}},
  \ and\ \bibinfo {author} {\bibfnamefont {E.~C.~G.}\ \bibnamefont
  {Sudarshan}},\ }\href {\doibase 10.1063/1.524655} {\bibfield  {journal}
  {\bibinfo  {journal} {Journal of Mathematical Physics}\ }\textbf {\bibinfo
  {volume} {21}},\ \bibinfo {pages} {2208} (\bibinfo {year} {1980})},\ \Eprint
  {http://arxiv.org/abs/https://doi.org/10.1063/1.524655}
  {https://doi.org/10.1063/1.524655} \BibitemShut {NoStop}%
\bibitem [{\citenamefont {Bohm}\ and\ \citenamefont {Gadella}(1989)}]{Bohm}%
  \BibitemOpen
  \bibfield  {author} {\bibinfo {author} {\bibfnamefont {A.}~\bibnamefont
  {Bohm}}\ and\ \bibinfo {author} {\bibfnamefont {M.}~\bibnamefont {Gadella}},\
  }\href@noop {} {\emph {\bibinfo {title} {Dirac Kets, Gamow Vectors and
  Gel’fand Triplets}}}\ (\bibinfo  {publisher} {Springer-Verlag Berlin
  Heidelberg},\ \bibinfo {year} {1989})\BibitemShut {NoStop}%
\bibitem [{\citenamefont {Chruscinski}(2003)}]{Chruscinski03}%
  \BibitemOpen
  \bibfield  {author} {\bibinfo {author} {\bibfnamefont {D.}~\bibnamefont
  {Chruscinski}},\ }\href@noop {} {\bibfield  {journal} {\bibinfo  {journal}
  {Journal of Mathematical Physics}\ }\textbf {\bibinfo {volume} {44}},\
  \bibinfo {pages} {3718} (\bibinfo {year} {2003})}\BibitemShut {NoStop}%
\bibitem [{\citenamefont {Chruscinski}(2004)}]{Chruscinski04}%
  \BibitemOpen
  \bibfield  {author} {\bibinfo {author} {\bibfnamefont {D.}~\bibnamefont
  {Chruscinski}},\ }\href@noop {} {\bibfield  {journal} {\bibinfo  {journal}
  {Journal of Mathematical Physics}\ }\textbf {\bibinfo {volume} {45}},\
  \bibinfo {pages} {841} (\bibinfo {year} {2004})},\ \Eprint
  {http://arxiv.org/abs/math-ph/0307047} {math-ph/0307047} \BibitemShut
  {NoStop}%
\bibitem [{\citenamefont {Datta}(1995)}]{Datta}%
  \BibitemOpen
  \bibfield  {author} {\bibinfo {author} {\bibfnamefont {S.}~\bibnamefont
  {Datta}},\ }\href@noop {} {\emph {\bibinfo {title} {Electronic transport in
  mesoscopic systems}}}\ (\bibinfo  {publisher} {Cambridge university press},\
  \bibinfo {year} {1995})\BibitemShut {NoStop}%
\bibitem [{\citenamefont {{Shimbori}}\ and\ \citenamefont
  {{Kobayashi}}(2000)}]{Shimbori00}%
  \BibitemOpen
  \bibfield  {author} {\bibinfo {author} {\bibfnamefont {T.}~\bibnamefont
  {{Shimbori}}}\ and\ \bibinfo {author} {\bibfnamefont {T.}~\bibnamefont
  {{Kobayashi}}},\ }\href@noop {} {\bibfield  {journal} {\bibinfo  {journal}
  {Nuovo Cimento B Serie}\ }\textbf {\bibinfo {volume} {115}},\ \bibinfo
  {pages} {325} (\bibinfo {year} {2000})},\ \Eprint
  {http://arxiv.org/abs/math-ph/9910009} {math-ph/9910009} \BibitemShut
  {NoStop}%
\bibitem [{\citenamefont {Isi}\ \emph {et~al.}(2019)\citenamefont {Isi},
  \citenamefont {Giesler}, \citenamefont {Farr}, \citenamefont {Scheel},\ and\
  \citenamefont {Teukolsky}}]{Isi19}%
  \BibitemOpen
  \bibfield  {author} {\bibinfo {author} {\bibfnamefont {M.}~\bibnamefont
  {Isi}}, \bibinfo {author} {\bibfnamefont {M.}~\bibnamefont {Giesler}},
  \bibinfo {author} {\bibfnamefont {W.~M.}\ \bibnamefont {Farr}}, \bibinfo
  {author} {\bibfnamefont {M.~A.}\ \bibnamefont {Scheel}}, \ and\ \bibinfo
  {author} {\bibfnamefont {S.~A.}\ \bibnamefont {Teukolsky}},\ }\href {\doibase
  10.1103/PhysRevLett.123.111102} {\bibfield  {journal} {\bibinfo  {journal}
  {Phys. Rev. Lett.}\ }\textbf {\bibinfo {volume} {123}},\ \bibinfo {pages}
  {111102} (\bibinfo {year} {2019})}\BibitemShut {NoStop}%
\bibitem [{\citenamefont {Chirenti}(2018)}]{Chirenti}%
  \BibitemOpen
  \bibfield  {author} {\bibinfo {author} {\bibfnamefont {C.}~\bibnamefont
  {Chirenti}},\ }\href {\doibase 10.1007/s13538-017-0543-7} {\bibfield
  {journal} {\bibinfo  {journal} {Brazilian Journal of Physics}\ }\textbf
  {\bibinfo {volume} {48}},\ \bibinfo {pages} {102} (\bibinfo {year}
  {2018})}\BibitemShut {NoStop}%
\bibitem [{\citenamefont {Abbott~et al}(2017)}]{LIGO1}%
  \BibitemOpen
  \bibfield  {author} {\bibinfo {author} {\bibfnamefont {B.~P.}\ \bibnamefont
  {Abbott~et al}} (\bibinfo {collaboration} {LIGO Scientific Collaboration and
  Virgo Collaboration}),\ }\href {\doibase 10.1103/PhysRevLett.119.161101}
  {\bibfield  {journal} {\bibinfo  {journal} {Phys. Rev. Lett.}\ }\textbf
  {\bibinfo {volume} {119}},\ \bibinfo {pages} {161101} (\bibinfo {year}
  {2017})}\BibitemShut {NoStop}%
\bibitem [{\citenamefont {McEuen}\ \emph {et~al.}(1990)\citenamefont {McEuen},
  \citenamefont {Alphenaar}, \citenamefont {Wheeler},\ and\ \citenamefont
  {Sacks}}]{McEuen}%
  \BibitemOpen
  \bibfield  {author} {\bibinfo {author} {\bibfnamefont {P.}~\bibnamefont
  {McEuen}}, \bibinfo {author} {\bibfnamefont {B.}~\bibnamefont {Alphenaar}},
  \bibinfo {author} {\bibfnamefont {R.}~\bibnamefont {Wheeler}}, \ and\
  \bibinfo {author} {\bibfnamefont {R.}~\bibnamefont {Sacks}},\ }\href
  {\doibase https://doi.org/10.1016/0039-6028(90)90896-G} {\bibfield  {journal}
  {\bibinfo  {journal} {Surface Science}\ }\textbf {\bibinfo {volume} {229}},\
  \bibinfo {pages} {312 } (\bibinfo {year} {1990})}\BibitemShut {NoStop}%
\bibitem [{\citenamefont {Comforti}\ \emph {et~al.}(2002)\citenamefont
  {Comforti}, \citenamefont {Chung}, \citenamefont {Heiblum}, \citenamefont
  {Umansky},\ and\ \citenamefont {Mahalu}}]{Heiblum}%
  \BibitemOpen
  \bibfield  {author} {\bibinfo {author} {\bibfnamefont {E.}~\bibnamefont
  {Comforti}}, \bibinfo {author} {\bibfnamefont {Y.~C.}\ \bibnamefont {Chung}},
  \bibinfo {author} {\bibfnamefont {M.}~\bibnamefont {Heiblum}}, \bibinfo
  {author} {\bibfnamefont {V.}~\bibnamefont {Umansky}}, \ and\ \bibinfo
  {author} {\bibfnamefont {D.}~\bibnamefont {Mahalu}},\ }\href@noop {}
  {\bibfield  {journal} {\bibinfo  {journal} {Nature}\ ,\ \bibinfo {pages}
  {515}} (\bibinfo {year} {2002})}\BibitemShut {NoStop}%
\bibitem [{\citenamefont {Floser}\ \emph {et~al.}(2010)\citenamefont {Floser},
  \citenamefont {Florens},\ and\ \citenamefont {Champel}}]{Floser}%
  \BibitemOpen
  \bibfield  {author} {\bibinfo {author} {\bibfnamefont {M.}~\bibnamefont
  {Floser}}, \bibinfo {author} {\bibfnamefont {S.}~\bibnamefont {Florens}}, \
  and\ \bibinfo {author} {\bibfnamefont {T.}~\bibnamefont {Champel}},\
  }\href@noop {} {\bibfield  {journal} {\bibinfo  {journal} {Phys. Rev. B}\
  }\textbf {\bibinfo {volume} {82}},\ \bibinfo {pages} {161408} (\bibinfo
  {year} {2010})}\BibitemShut {NoStop}%
\bibitem [{\citenamefont {Aguirregabiria}\ and\ \citenamefont
  {Vishveshwara}(1996)}]{Aguirregabiria}%
  \BibitemOpen
  \bibfield  {author} {\bibinfo {author} {\bibfnamefont {J.}~\bibnamefont
  {Aguirregabiria}}\ and\ \bibinfo {author} {\bibfnamefont {C.}~\bibnamefont
  {Vishveshwara}},\ }\href {\doibase
  https://doi.org/10.1016/0375-9601(95)00937-X} {\bibfield  {journal} {\bibinfo
   {journal} {Physics Letters A}\ }\textbf {\bibinfo {volume} {210}},\ \bibinfo
  {pages} {251 } (\bibinfo {year} {1996})}\BibitemShut {NoStop}%
\bibitem [{\citenamefont {{Maldacena}}\ \emph {et~al.}(2016)\citenamefont
  {{Maldacena}}, \citenamefont {{Shenker}},\ and\ \citenamefont
  {{Stanford}}}]{Maldacena16}%
  \BibitemOpen
  \bibfield  {author} {\bibinfo {author} {\bibfnamefont {J.}~\bibnamefont
  {{Maldacena}}}, \bibinfo {author} {\bibfnamefont {S.~H.}\ \bibnamefont
  {{Shenker}}}, \ and\ \bibinfo {author} {\bibfnamefont {D.}~\bibnamefont
  {{Stanford}}},\ }\href {\doibase 10.1007/JHEP08(2016)106} {\bibfield
  {journal} {\bibinfo  {journal} {Journal of High Energy Physics}\ }\textbf
  {\bibinfo {volume} {8}},\ \bibinfo {eid} {106} (\bibinfo {year} {2016})},\
  \Eprint {http://arxiv.org/abs/1503.01409} {arXiv:1503.01409 [hep-th]}
  \BibitemShut {NoStop}%
\bibitem [{\citenamefont {Shenker}\ and\ \citenamefont
  {Stanford}(2014)}]{Shenker14}%
  \BibitemOpen
  \bibfield  {author} {\bibinfo {author} {\bibfnamefont {S.}~\bibnamefont
  {Shenker}}\ and\ \bibinfo {author} {\bibfnamefont {D.}~\bibnamefont
  {Stanford}},\ }\href {\doibase 10.1007/JHEP03(2014)067} {\bibfield  {journal}
  {\bibinfo  {journal} {Journal of High Energy Physics}\ }\textbf {\bibinfo
  {volume} {3}},\ \bibinfo {eid} {67} (\bibinfo {year} {2014})}\BibitemShut
  {NoStop}%
\bibitem [{\citenamefont {Hashimoto}\ and\ \citenamefont
  {Tanahashi}(2017)}]{Hashimoto}%
  \BibitemOpen
  \bibfield  {author} {\bibinfo {author} {\bibfnamefont {K.}~\bibnamefont
  {Hashimoto}}\ and\ \bibinfo {author} {\bibfnamefont {N.}~\bibnamefont
  {Tanahashi}},\ }\href {\doibase 10.1103/PhysRevD.95.024007} {\bibfield
  {journal} {\bibinfo  {journal} {Phys. Rev. D}\ }\textbf {\bibinfo {volume}
  {95}},\ \bibinfo {pages} {024007} (\bibinfo {year} {2017})}\BibitemShut
  {NoStop}%
\bibitem [{\citenamefont {Morita}(2018)}]{Morita18}%
  \BibitemOpen
  \bibfield  {author} {\bibinfo {author} {\bibfnamefont {T.}~\bibnamefont
  {Morita}},\ }\href@noop {} {\  (\bibinfo {year} {2018})},\ \Eprint
  {http://arxiv.org/abs/1801.00967} {arXiv:1801.00967} \BibitemShut {NoStop}%
\end{thebibliography}%

\end{document}